\documentclass[]{article}

\usepackage{icrc2011}

\title{VERITAS observations of the Segue 1 dwarf spheroidal galaxy}

\newcommand{\etal}{\MakeLowercase{\textit{et al. }}} 
\shorttitle{M. Vivier \etal VERITAS observations of the Segue 1 dsph}

\authors{Matthieu Vivier$^{1}$ for the VERITAS collaboration$^{2}$}
\afiliations{$^1$Bartol Research Institute, Department of Physics and Astronomy, University of Delaware, Newark DE 19716\\ $^2$see J. Holder et al. (these proceedings) or http://veritas.sao.arizona.edu/conferences/authors?icrc2011}
\email{mvivier@bartol.udel.edu}

\abstract{In the cosmological paradigm, Cold Dark Matter (DM) dominates the mass content of 
the Universe and is present at every scale. Candidates for DM include many 
extensions of the standard model, with a Weakly Interacting Massive Particle (WIMP) 
in the mass range from 50 GeV to greater than 10 TeV. The self-annihilation of 
WIMPs in astrophysical regions of high DM density can produce secondary particles 
including Very High Energy (VHE) gamma rays with energies up to the DM particle 
mass. The VERITAS array of Cherenkov telescopes, designed for the detection of VHE 
gamma rays in the 100 GeV-10 TeV energy range, is an appropriate instrument for 
the detection of DM. Among the possible astrophysical targets, dwarf spheroidal 
galaxies (dSphs) of the Local Group are promising targets to search for the 
annihilation signature of DM due to their proximity and large DM content. We report here on extensive observations conducted by VERITAS on the 
nearby Segue 1 satellite galaxy, which is currently considered as one of the best 
dSphs for DM studies. The results are discussed in the framework of WIMP models, 
with a special emphasis on leptophilic DM models invoked to explain the recent 
cosmic-ray lepton anomalies.}

\begin{document}
\maketitle

\section{Introduction}
The compelling evidences for the presence of Dark Matter (DM) in the different structures of the Universe \cite{DM} have motivated numerous efforts to search for DM by means of astrophysical observations. If DM is made of Weakly Interacting Massive Particles (WIMPs) annihilating into standard model particles, indirect searches for DM annihilations with very high energy (VHE) $\gamma$-rays provide one of the best way to constrain the nature of the DM particle: $\gamma$-rays are free of any propagation effects on short distances ($\mathrm{\leq}$ 1\,Mpc) and DM particle annihilation is predicted to give a unique $\gamma$-ray spectrum. Such searches are often conducted using pointed observations toward nearby DM overdensities, because the annihilation rate is proportional to the squared DM density. Popular targets include the Galactic Center \cite{GCBuckley, GCWhipple,GCDM1,GCDM2}, satellite galaxies of the Milky-Way (MW)  \cite{dSphDM, dSph1,dSph2,dSph3,dSph4,dSph5,MAGICSegue1}, globular clusters \cite{GloC1,GloC2} and clusters of galaxies \cite{Cluster}.\\
The dwarf spheroidal galaxies (dSphs) of the Local Group best meet the criteria for a clear and unambiguous detection of DM. They are gravitationally bound objects and contain up to $\cal{O}$($\mathrm{10^{3}}$) more mass in DM than in visible matter. As opposed to the Galactic Center, they are environments with a favorably low astrophysical $\gamma$-ray background. Neither astrophysical $\gamma$-ray sources (supernova remnants, pulsar wind nebulae, etc) nor gas acting as target material for cosmic rays have been observed in these systems \cite{dSph}. Furthermore, their relative proximity and high galactic latitude make them the best astrophysical targets for high signal-to-noise detection.\\
Segue 1 is an ultra faint dSph discovered in 2006 as an overdensity of resolved stars in the Sloan Digital Sky Survey \cite{Segue1SDSS}. It is located at a distance of 23 $\mathrm{\pm}$ 2 kpc from the Sun at (Ra,Dec) = (10$\mathrm{^{h}}$07$\mathrm{^{m}}$03.2$\mathrm{^{s}}$,16d04'25''), well above the Galactic plane. Because of its proximity to the Sagittarius stream, the nature of Segue 1 overdensity has recently been disputed, some people arguing that it was a tidally disrupting star cluster originally associated with the Sagittarius dSph \cite{Segue1GloC}. However, a recent kinematic study of a larger member star sample (66 stars compared to the previous 24 stars sample) has firmly confirmed that Segue 1 is an ultra faint MW satellite galaxy \cite{Segue1dSph}. According to the study of its star kinematics, Segue 1 is the most DM-dominated dSph and is often highlighted as the best dSph target for indirect DM searches \cite{Segue1HL1,Segue1HL2,Segue1HL3}.
\section{Observations}
VERITAS \cite{VERITAS} is an array of four 12 meter imaging atmospheric Cherenkov telescopes (IACTs) located at the Fred Lawrence Whipple Observatory (FLWO) in southern Arizona. Each telescope is composed of a large mirror area which reflects Cherenkov light induced by particle air showers on a camera placed in the focal plane. Each of the VERITAS cameras consists of 499 photomultiplier tubes covering a 3.5$^{\circ}$ field of view. The large collection area ($\sim$10$^{5}$ m$^{2}$), in conjunction with the stereoscopic imaging of air showers, allow VERITAS to detect VHE gamma-rays between energies of 100 GeV and 30 TeV with an energy and angular resolution of $\sim$15-25\% and $\sim$ 0.1$^{\circ}$, respectively. A source with a 1\% Crab Nebula flux can be detected by VERITAS in approximately 25 hours.\\
Observations of the Segue 1 dSph were performed between January 2010 and May 2011, at a mean zenith angle of $\mathrm{\sim 20^{\circ}}$. Only data taken under good weather conditions and with no major hardware problems were selected. The total exposure, after dead time correction, is nearly 48 hours. This is the biggest exposure reported so far by any IACT array on a dSph. The observations have been conducted using the "wobble" pointing mode, where the camera center is offset by 0.5$\mathrm{^{\circ}}$ from the target position. This allows for simultaneous background estimation and source observation, thus reducing the systematic uncertainties in the background determination. Four wobble directions were used (North, South, East, West) and were alternated from run to run.
\begin{figure*}[!t]
   \centerline{\includegraphics[width=2.8in]{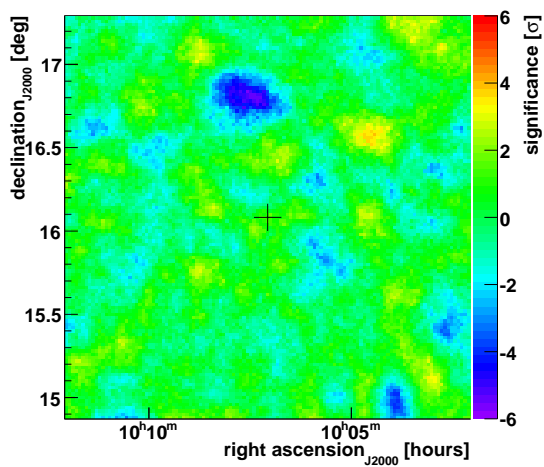}}
   \caption{Significance map obtained after $\gamma$-ray selection and background subtraction. The significance is calculated according to the Li \& Ma method \cite{LiMa}. The black cross indicates the target position.}
   \label{fig1}
 \end{figure*}
\section{Data analysis and results}
Data reduction follows the methods described in \cite{VERITASdataanalysis}. After calibration of the camera gains, images recorded by each of the VERITAS telescope are characterized by a second moment analysis giving the Hillas parameters. A stereoscopic analysis combining each telescope image parameters is used to reconstruct the $\gamma$-ray arrival direction and shower core position. Cuts on the Hillas parameters and the angle $\mathrm{\theta}$ between the target position and the reconstructed arrival direction ($\mathrm{\theta^{2}}$ $\mathrm{\leq}$ 0.015 deg$\mathrm{^{2}}$, defining the source region) are applied to reduce the cosmic-ray background. The image distance from the center of the camera is also required to be less than 2$\mathrm{^{\circ}}$ to avoid truncation effects. Furthermore, the integrated charge recorded in at least two telescopes is required to be $\mathrm{\geq}$ 90 photoelectrons which effectively sets the analysis energy threshold\footnote{The energy threshold quoted here is taken to be the energy at which the differential detection rate of $\gamma$-rays from the Crab Nebula peaks.} to be 170 GeV.  
\\
After the $\gamma$-ray selection, the residual background was estimated using the ring background technique \cite{RBck}. Two circular regions, of radius $\mathrm{0.2^{\circ}}$ centered on the target position, and of radius $\mathrm{0.3^{\circ}}$ centered on the bright star $\eta$-Leonis (with apparent magnitude in the visible band V = 3.5, and located $\mathrm{0.68^{\circ}}$ from the position of Segue 1), were excluded for the background determination. The resulting significance map is shown on figure \ref{fig1}. No significant $\gamma$-ray excess is found at any positions within the field of view. The large depletion area, with negative significances, corresponds to $\mathrm{\eta}$-Leonis. Given the absence of signal, one can derive an upper limit (UL) on the number of $\gamma$-rays in the source region. Because of the energy reconstruction bias in the low energy regime near the threshold, we define a minimum energy $\mathrm{E_{min}=300\,GeV}$ to safely compute ULs. Using the Rolke method \cite{RolkeMethod}, the UL at the 95\% confidence level (CL) in the number of excess events in the source region is:
\begin{equation}
\mathrm{N_{\gamma}^{95\% CL}(E>300\,GeV) = 102.5}\,,
\end{equation}
which gives an integral flux UL above 300 GeV of, assuming a power law spectrum with index $\mathrm{\Gamma\,=\,2.6}$:
\begin{equation}
\mathrm{\Phi_{\gamma}^{95\% CL}(E>300\,GeV) < 8 \times 10^{-13}\,cm^{-2}\,s^{-1}}
\end{equation}
This corresponds to 0.5\% of the Crab Nebula integral flux.
\section{Dark Matter annihilation bounds}
The absence of signal at the position of Segue 1 can be used to derive constraints on various DM models. The $\gamma$-ray flux from the annihilations of DM particles, of mass $\mathrm{m_{DM}}$, in a spherical DM halo is given by a particle physics term times an astrophysics term:
\begin{equation}
\mathrm{\frac{d\Phi_{\gamma}}{dE}(\Delta\Omega,E)=\frac{\langle\sigma v \rangle}{8\,\pi\,m_{DM}^{2}}\,\frac{dN_{\gamma}}{dE}\times\bar{J}(\Delta\Omega)},
\end{equation}
where the astrophysical factor $\mathrm{\bar{J}(\Delta\Omega)}$ is the squared DM density integrated along the line of sight (los) and over the solid angle $\mathrm{\Delta\Omega}$. The solid angle is given here by the size of the signal search region defined previously in our analysis, i.e. $\mathrm{\theta^{2}}$ $\mathrm{\leq}$ 0.015 deg$\mathrm{^{2}}$. The particle physics term contains all the information about the DM particle: its mass $\mathrm{m_{DM}}$, its total velocity-weighted annihilation cross-section $\mathrm{\langle\sigma v \rangle}$ and the differential $\gamma$-ray spectrum from all final states weighted by their corresponding branching ratios, $\mathrm{dN_{\gamma}/dE}$.\\
The estimate of the astrophysical factor requires a modeling of the Segue 1 DM profile. An Einasto profile \cite{Einasto} is used:
\begin{equation}
\mathrm{\rho_{DM}(r) = \rho_s\,e^{-2n\,[(r/r_s)^{1/n}-1]}}, 
\end{equation}
with the scale density, the scale radius and the index n respectively being $\mathrm{\rho_s= 1.1\times 10^{8}\,M_{\odot}\,kpc^{-3}}$, $\mathrm{r_s = 0.15\,kpc}$ and n = 3.3 \cite{MAGICSegue1,Private}. The value of the tidal radius changes the astrophysical factor by less than 10\%. We adopt a value of 500 pc, which is the median truncation radius of the Via Lactea II simulation subhalos presenting similar characteristics to the Segue 1 DM halo \cite{Segue1dSph,Clumps3}. Having these parameters in hand, the value of the astrophysical factor within the solid angle subtended by our source region is $\mathrm{\bar{J}(\Delta\Omega) = 7.7\times 10^{18}\,GeV^{-2}\,cm^{-5}\,sr}$. The systematic uncertainties quoted in the latest modeling of the Segue 1 DM distribution are less than an order of magnitude \cite{Segue1HL1}.\\
Once the astrophysical factor has been estimated, the UL in the number of $\gamma$-ray candidates in the source region can be translated into an UL on the total annihilation cross-section:
\begin{equation}
\mathrm{\langle\sigma v\rangle_{min}^{95\%\,CL} = \frac{8\pi}{\bar{J}(\Delta\Omega)}\times
\frac{N_{\gamma}^{95\% CL}m_{DM}^{2}}{T_{obs}\,\int_0^{m_{DM}}{\cal{A}}_{eff}(E)\frac{dN_{\gamma}}{dE}dE}},
\end{equation}
where $\mathrm{T_{obs}}$ is the total observation time and $\mathrm{{\cal{A}}_{eff}(E)}$ is the effective area as a function of energy, zenith angle and the offset of the source from the pointing position.
\subsection{Classical constraints}
\begin{figure}[!ht]
  \centering
  \includegraphics[width=3in]{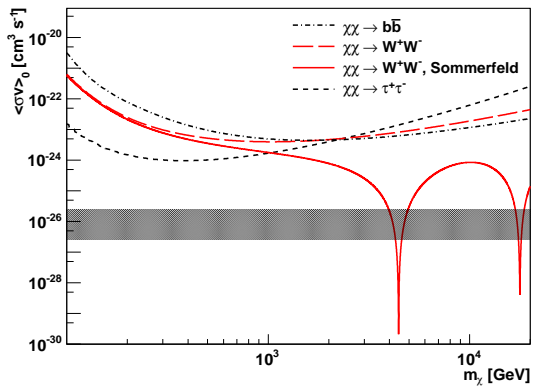}
  \caption{Upper limits at the 95\% CL on $\mathrm{\langle\sigma v\rangle_0}$ as a function of the DM particle mass for different "classical" annihilation channels. The grey band area represents a range of generic values for the annihilation cross-section in the case of thermally produced DM. The red solid curve corresponds to the Sommerfeld enhancement effect that could possibly arises for the $\mathrm{ W^{+}W^{-}}$ channel (see text for details).}
  \label{fig2}
 \end{figure}

Independently of the DM model, we consider in this section "classical" annihilation channels, each time assuming a 100\% branching ratio. Figure \ref{fig2} shows our constraints on $\mathrm{\langle\sigma v\rangle}$ as a function of the DM particle mass for three annihilation channels $\mathrm{\chi\chi\rightarrow\tau^{+}\tau^{-}}$, $\mathrm{\chi\chi\rightarrow W^{+}W^{-}}$ and $\mathrm{\chi\chi\rightarrow b\bar{b}}$. The $\gamma$-ray spectra have been simulated with the particle physics event generator PYTHIA v8.1 \cite{PYTHIA}. The $\mathrm{b\bar{b}}$ and $\mathrm{\tau^{+}\tau^{-}}$ channels encompass all possible DM $\gamma$-ray annihilation spectra in the VERITAS energy range, and give an idea of the uncertainties related to the particle physics model. For the $\mathrm{W^{+}W^{-}}$ channel, the 95\% CL UL on $\mathrm{\langle\sigma v\rangle}$ is at the level of $\mathrm{8\times10^{-24}\,cm^{3}\,s^{-1}}$ around 1 TeV, among the best reported so far with dSph observations.
\\
We also consider the case where the annihilation cross-section can be boosted with the {\it Sommerfeld} enhancement effect. The Sommerfeld effect arises when the two DM particles interact through an attractive potential, mediated by the exchange of a massive boson \cite{SommerfeldW}. The annihilation cross-section enhancement is particularly effective when the DM particle relative velocity is very small. Depending on the mass $\mathrm{m_{\phi}}$ and the coupling $\mathrm{\alpha}$ of the exchanged boson, the Sommerfeld enhancement can exhibit a serie of resonances for specific values of the DM particle mass, giving very large boost factors up to $\mathrm{10^{6}}$. In such scenarios, the annihilation cross-section is given by:
\begin{equation}
\mathrm{\langle\sigma v\rangle=\bar{S}\times\langle\sigma v\rangle_0},
\end{equation}
where $\mathrm{\bar{S}}$ is the Sommerfeld boost factor averaged over the DM particle velocity distribution\footnote{In the case of Segue 1, we assume that the relative velocity distribution is Maxwellian (i.e. the DM gaz is thermalized and in equilibrium), with a width of the order of the mean star velocity dispersion. See \cite{Segue1HL1} for further details.} and depends on the DM particle mass. $\mathrm{\langle\sigma v\rangle_0}$ is the thermal WIMP annihilation cross-section before freeze-out. The Sommerfeld enhancement is of particular interest for cold DM halos like dSphs, where the mean star velocity dispersion can reach $\mathrm{10^{-5}\,c}$. Figure \ref{fig2} shows the limits on $\mathrm{\langle\sigma v\rangle_0}$, accounting for the corresponding Sommerfeld boost in the case of DM annihilating into a pair of $\mathrm{W^{+}W^{-}}$ bosons when a $\mathrm{Z^{0}}$ boson ($\mathrm{m_{Z^{0}}\sim 90\,GeV}$, $\mathrm{\alpha \sim 1/30}$) is exchanged. The exclusion curve exhibits resonances, which correspond to the first resonances in the Sommerfeld enhancement for DM particle masses of 4.5 and 17 TeV, respectively. VERITAS severely excludes these first resonances.
\subsection{Models with a new force in the dark sector}
\begin{figure}[!ht]
  \centering
  \includegraphics[width=3in]{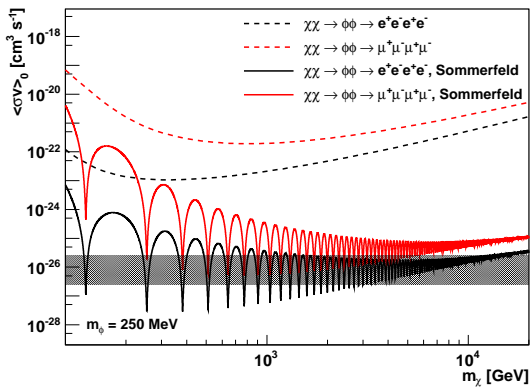}
  \caption{95\% CL upper limits on $\mathrm{\langle\sigma v\rangle_0}$ as a function of the DM particle mass for lepton annihilation channels, in the framework of particle physics models with a new force in the dark sector. The greay band area in the same as in figure \ref{fig2}. The dashed curves correspond to limits without any enhancement to $\mathrm{\langle\sigma v\rangle_0}$. The solid curves includes the Sommerfeld effect expected within these models.}
  \label{fig3}
 \end{figure}

The last section is dedicated to a class of "leptophilic" DM models invoked to explain the recent cosmic-ray lepton anomalies measured by ATIC \cite{ATIC} and PAMELA \cite{PAMELA}. In any DM interpretation, the cosmic-ray lepton excesses require a DM particle mostly annihilating into leptons and with a high annihilation cross-section (compared to $\mathrm{\langle\sigma v\rangle_0 \sim 3\times10^{-26}\,cm^{3}\,s^{-1}}$). Models which introduce a new force in the dark sector have recently been proposed to fit these requirements \cite{AH}. The new force is carried by a light scalar field $\mathrm{\phi}$ predominantly decaying into leptons and with a mass and coupling to standard model particles chosen to prevent the overproduction of antiprotons. These models naturally comprise a Sommerfeld enhancement to the annihilation cross-section. As for the classic $\mathrm{W^{+}W^{-}}$ channel, the boost also exhibits a serie of resonances, but for different DM particle masses because the mass of the exchanged boson and its coupling constant are different\footnote{The value of the coupling constant $\mathrm{\alpha}$ is fixed here by the computation of the DM relic density assuming that the annihilation process $\mathrm{\chi\chi\rightarrow\phi\phi}$ is the only channel that regulates the DM density before freeze-out. See \cite{SommerfeldRelic} for further details.}.\\
Figure \ref{fig3} shows the constraints on such models, for the two leptonic annihilation modes $\mathrm{\chi\chi\rightarrow\phi\phi\rightarrow e^{+}e^{-}e^{+}e^{-}}$ and $\mathrm{\chi\chi\rightarrow\phi\phi\rightarrow \mu^{+}\mu^{-}\mu^{+}\mu^{-}}$, and for a scalar particle $\mathrm{\phi}$ of mass 250 MeV. VERITAS excludes most of the resonances, especially for the $\mathrm{e^{+}/e^{-}}$ channel. This result holds and strengthens for larger $\mathrm{\phi}$ masses, thus disfavoring this class of models.
\section*{Acknowledgements}
We thank Rouven Essig, Neelima Sehgal and Louis E. Strigari for useful discussions about the Segue 1 dark matter distribution profile. This research is supported by grants from the US Department of Energy, the US National Science Foundation, and the Smithsonian Institution, by NSERC in Canada, by Science Foundation Ireland, and by STFC in the UK. We acknowledge the excellent work of the technical support staff at the FLWO and at the collaborating institutions in the construction and operation of the instrument.

\clearpage


\begin{thebibliography}{}

\bibitem{DM} Bertone G., et al., Phys. Rep., {\bf 405}, 279 (2005)
\bibitem{GCBuckley} Bergstrom L., et al., Astropart. Phys., {\bf 9}, 137 (1998)
\bibitem{GCWhipple} Kosack K., et al., ApJL, {\bf 608}, 97 (2004)
\bibitem{GCDM1} Aharonian F., et al., Phys. Rev. Lett., {\bf 97}, 221102 (2006)
\bibitem{GCDM2} Abramowski A., et al., Phys. Rev. Lett., {\bf 106}, 161301 (2011)
\bibitem{dSphDM} Evans N.W., et al., Phys. Rev. D, {\bf 69}, 123501 (2004)
\bibitem{dSph1} Acciari V. A., et al., ApJ, {\bf 720}, 1174 (2010)
\bibitem{dSph2} Aharonian F., et al., Astropart. Phys., {\bf 29}, 55 (2008)
\bibitem{dSph3} Aharonian F., et al., ApJ, {\bf 691}, 175 (2009)
\bibitem{dSph4} Abramowski A., et al., Astropart. Phys., {\bf 34}, 608 (2011)
\bibitem{dSph5} Aliu E., et al., ApJ, {\bf 697}, 1299 (2009)
\bibitem{MAGICSegue1} J. Aleksic, et al., JCAP, {\bf 06}, 035, (2011)
\bibitem{GloC1} Wood M., et al.,  ApJ, {\bf 678}, 594 (2008)
\bibitem{GloC2} Abramowski A., et al., ApJ, {\bf 735}, 12 (2011)
\bibitem{Cluster} Aleksic J., et al., ApJ, {\bf 710}, 634 (2010)
\bibitem{dSph} Mateo M.,  Annu. Rev. Astron. Astrophys., {\bf 36}, 435 (1998)
\bibitem{Segue1SDSS} Belokurov V., et al., ApJ, {\bf 654}, 897 (2007)
\bibitem{Segue1GloC} Niederste-Ostholt M., et al., MNRAS, {\bf 398}, 1771 (2009)
\bibitem{Segue1dSph} Simon J. D., et al., ApJ, {\bf 733}, 46 (2011)
\bibitem{Clumps3} J. Diemand, et al., Nature, {\bf 454}, 735 (2008)
\bibitem{Segue1HL1} Martinez G. D., et al., JCAP, {\bf 0906}, 014 (2009)
\bibitem{Segue1HL2} Essig R., et al., Phys. Rev. D, {\bf 80}, 023506 (2009)
\bibitem{Segue1HL3} Essig R., et al., Phys. Rev. D, {\bf 82}, 123503 (2010)
\bibitem{VERITAS} Holder J. et al., these proceedings (2011)
\bibitem{VERITASdataanalysis} Acciari V. A., et al., ApJ, {\bf 679}, 1427 (2008)
\bibitem{RBck} Berge D., et al., A\&A, {\bf 466}, 1219 (2007)
\bibitem{LiMa} Li T. and Ma Y., ApJ, {\bf 272}, 317 (1983)
\bibitem{RolkeMethod} Rolke W.A., et al., Nucl. Instrum. Methods, {\bf A551}, 493 (2005)
\bibitem{Einasto} Navarro J. F., et al, MNRAS, {\bf 402}, 21 (2010)
\bibitem{Private} R. Essig, N. Sehgal \& L.E. Strigari, Private communication
\bibitem{PYTHIA} http://home.thep.lu.se/~torbjorn/Pythia.html
\bibitem{SommerfeldW} Lattanzi M. and Silk J., Phys. Rev. D, {\bf 79}, 083523 (2009)
\bibitem{ATIC} Chang J., et al., Nature, {\bf 456}, 362 (2008)
\bibitem{PAMELA} Adriani O., et al, Nature, {\bf 458}, 607 (2009)
\bibitem{AH} Arkani-Hamed N., et al., Phys. Rev. D, {\bf 79}, 015014 (2009)
\bibitem{SommerfeldRelic} Feng J., et al., Phys. Rev. D, {\bf 82}, 083525 (2010)
\end{thebibliography}
\end{document}